\documentclass[conference, 9pt]{IEEEtran}
\IEEEoverridecommandlockouts

\usepackage{cite}
\usepackage{enumitem}
\usepackage{amsmath,amssymb,amsfonts}
\usepackage{algorithm}
\usepackage{algpseudocode}

\usepackage{setspace}

\usepackage{booktabs}

\usepackage{graphicx}
\usepackage{textcomp}
\usepackage{xcolor}
\usepackage{wrapfig}

\usepackage{stfloats}

\usepackage{caption}
\usepackage[a4paper, total={184mm,239mm}]{geometry}
\def\BibTeX{{\rm B\kern-.05em{\sc i\kern-.025em b}\kern-.08em
    T\kern-.1667em\lower.7ex\hbox{E}\kern-.125emX}}
\begin{document}

\title{CognitiveArm: Enabling Real-Time EEG-Controlled Prosthetic Arm Using Embodied Machine Learning
}

\author{
\IEEEauthorblockN{Abdul Basit, Maha Nawaz, Saim Rehman, Muhammad Shafique}
\IEEEauthorblockA{\textit{eBRAIN Lab, Division of Engineering}
\textit{New York University (NYU) Abu Dhabi}
Abu Dhabi, UAE \\
\{abdul.basit, mzn2386, sr7849, muhammad.shafique\}@nyu.edu }
\vspace{-2em}
}
\maketitle

\begin{abstract}

Efficient control of prosthetic limbs via non-invasive brain-computer interfaces (BCIs) requires advanced EEG processing, including pre-filtering, feature extraction, and action prediction, performed in real time on edge AI hardware. Achieving this on resource-constrained devices presents challenges in balancing model complexity, computational efficiency, and latency. We present CognitiveArm, an EEG-driven, brain-controlled prosthetic system implemented on embedded AI hardware, achieving real-time operation without compromising accuracy. The system integrates BrainFlow, an open-source library for EEG data acquisition and streaming, with optimized deep learning (DL) models for precise brain signal classification. Using evolutionary search, we identify Pareto-optimal DL configurations through hyperparameter tuning, optimizer analysis, and window selection, analyzed individually and in ensemble configurations. We apply model compression techniques such as pruning and quantization to optimize models for embedded deployment, balancing efficiency and accuracy. We collected an EEG dataset and designed an annotation pipeline enabling precise labeling of brain signals corresponding to specific intended actions, forming the basis for training our optimized DL models. CognitiveArm also supports voice commands for seamless mode switching, enabling control of the prosthetic arm's 3 degrees of freedom (DoF). Running entirely on embedded hardware, it ensures low latency and real-time responsiveness. A full-scale prototype, interfaced with the OpenBCI UltraCortex Mark IV EEG headset, achieved up to 90\% accuracy in classifying three core actions (left, right, idle). Voice integration enables multiplexed, variable movement for everyday tasks (e.g., handshake, cup picking), enhancing real-world performance and demonstrating CognitiveArm's potential for advanced prosthetic control.

\end{abstract}

\begin{spacing}{0.99}

\section{Introduction}

\begin{figure*}[b]
    \centering
    \includegraphics[width=\linewidth]{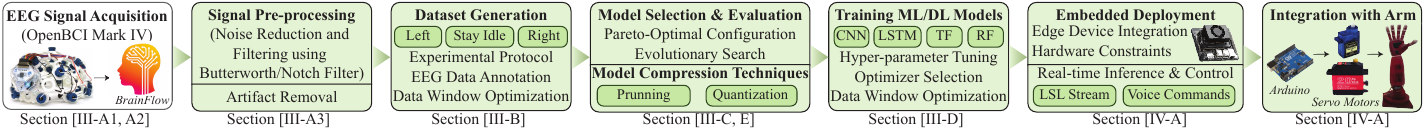}
    \caption{Overview of the CognitiveArm system pipeline, including input acquisition, dataset generation, and Pareto-optimal model selection, and actuation.}
    \label{fig:intrduction}
\end{figure*}

The integration of non-invasive brain-computer interfaces (BCIs) with prosthetic limbs holds significant potential for enhancing assistive technologies, particularly for individuals with upper limb amputations or paralysis. Traditional control methods, such as surface electromyography (sEMG), rely on residual muscle activity, which can be unreliable or absent in conditions like amyotrophic lateral sclerosis (ALS), spinal cord injuries, and brainstem strokes~\cite{fitzgibbons2015functional}. In such cases, sEMG-based systems often fail to provide effective control due to muscle atrophy and ankylosis (see Table~\ref{tab:diseases}). Electroencephalography (EEG)-based BCIs offer a promising alternative by directly interpreting brain signals, enabling prosthetic control even in the absence of functional muscles~\cite{Padfield2022,Wang2019}.

\begin{table}[h!]
\centering
\renewcommand{\arraystretch}{1} 
\setlength{\tabcolsep}{2pt} 
\scriptsize
\begin{tabular}{|p{1.4cm}|p{3.6cm}|p{3.5cm}|}
\hline
\textbf{Condition} & \textbf{Impact on EMG Use} & \textbf{EEG as a Solution} \\ \hline
ALS & Muscle atrophy limits residual EMG signals & EEG-based BCI can interpret brain signals directly \cite{Ma2024, trial} \\ \hline
Spinal Cord Injury & Loss of voluntary muscle control below the injury & EEG can bypass muscle control pathways \cite{Leng2024, Li2023} \\ \hline
Brainstem Stroke & Severe loss of motor control, leading to locked-in syndrome & EEG can control assistive devices using brain signals \cite{Maslova2023} \\ \hline
Multiple Sclerosis & Muscle spasticity and weakness reduce EMG effectiveness & EEG can offer more reliable control options \cite{trial} \\ \hline
Muscular Dystrophies & Progressive muscle degeneration limits EMG utility & EEG allows control through brain signals \cite{Maslova2023, Ma2024} \\ \hline
\end{tabular}
\caption{Comparison of EMG and EEG effectiveness in various conditions}
\label{tab:diseases}
\end{table}

\textbf{The research problem targeted in this paper} is to design an intelligent prosthetic arm with real-time embodied ML functions for multi-action EEG classification and voice command integration, which require developing efficient DL models and low-cost prototype, while ensuring low-latency and high accuracy under stringent design constraints.


\textbf{State-of-the-Art Solutions and Their Limitation}: Current state-of-the-art systems often rely on powerful personal computers or cloud-based servers for real-time EEG processing, which limits portability and introduces latency unsuitable for prosthetic control~\cite{s21134293}. While DL models such as convolutional neural networks (CNNs), long short-term memory (LSTM) networks, and transformers have shown promise in extracting temporal and spatial features from EEG signals~\cite{app14146347,Samal,Craik2019-mi,Roy_2019}, these prior works often suffer from several limitations:

\begin{itemize}
    \item \underline{\textit{Lack of Full System Integration}}: Many existing works focus on EEG signal classification in isolation and are not integrated into functional prosthetic prototypes with limited actions  ~\cite{app14146347,Samal,Craik2019-mi,Roy_2019}. 
    \item \underline{\textit{Computational Demands}}: The complexity of DL models can hinder deployment on embedded devices due to limited computational resources and memory constraints \cite{7994570}. 
    \item \underline{\textit{Latency and Real-Time Performance}}: Timing and latency, critical factors for real-time applications, are not thoroughly investigated. Solutions utilizing cloud-based processing introduce unacceptable latency and raise privacy concerns due to the transmission of sensitive neural data \cite{pandemic}.
    \item \underline{\textit{Model Compression Trade-offs}}: Techniques like model quantization and pruning have been explored to reduce computational load \cite{7994570}, but they often come at the cost of accuracy, which is crucial for precise prosthetic control \cite{9892902, 9211496}.
\end{itemize}

Our research focuses on developing an intelligent prosthetic arm system that integrates high-performance EEG signal processing with embedded DL to ensure real-time functionality. 
The system's core is a low-latency DL-based action prediction framework that predicts \textit{Action Labels}, which correspond to the user's intention among three core actions which lead to a variable amount of change in the position of the arm, enabling precise low-level control while adhering to resource constraints. 
These core actions are used to perform different tasks such as grabbing objects, handshake, etc. 
\textit{Noteworthy, full system integration is crucial to validate the efficiency of the DL-driven control loop, encompassing sensing, decision-making, motor control, and actuation in a synergistic way, enabling seamless real-world operation.}

\textbf{Research Challenges}: These observations expose several research challenges and requirements (outlined below) in devising solutions for the targeted research problem. See overview in Figure 1.
\begin{itemize} 
    \item \underline{\textit{Computational Complexity}}: Developing DL models that are both accurate and efficient enough to run in real-time on resource-constrained edge devices.
    \item \underline{\textit{Accuracy vs. Efficiency}}: Investigating the impact of model compression techniques on accuracy and latency to ensure precise control without sacrificing significant performance.
    \item \underline{\textit{EEG Dataset Generation and Annotation}}: Creating a robust dataset generation and annotation pipeline to accurately capture and label EEG signals corresponding to specific intended actions.
    \item \underline{\textit{Cost-effective Design}}: Prototyping affordable prosthetic arms while ensuring functionality and robustness remains a key challenge, particularly for developing regions in the World.
    \item \underline{\textit{Latency}}: Ensuring low latency in the end-to-end system to facilitate real-time interaction and control.


\end{itemize}
\textbf{Our Novel Contributions}: To address these challenges, we introduce \textit{\textbf{CognitiveArm}} that integrates non-invasive real-time EEG processing and classification using Embodied Learning with advanced DL architectures on edge AI processors for control of prosthetic limbs. To enable this, our CognitiveArm system employs the following:
\begin{itemize} 

    \item\underline{\textit{An On-Device DL Engine}} that leverages CNN, LSTM, Random Forest, and Transformer models individually and in an ensemble configuration for action prediction. This approach optimizes temporal and spatial feature extraction, ensuring high accuracy in EEG signal classification.

    \item\underline{\textit{Efficient DL Model Design Space Exploration Method}} employing an evolutionary search algorithm for identifying Pareto-optimal DL model configurations while balancing accuracy and efficiency trade-off. This process involves hyper-parameter tuning, optimizer selection, and window size selection to achieve the best trade-off between computational load and precision.

    \item\underline{\textit{EEG Dataset Generation and Annotation Pipeline}} to enable EEG data collection and a labeling protocol, ensuring precise and standardized annotation of brain signals corresponding to specific intended actions (left, right, stay idle) for each user.
    \item\underline{\textit{Model Compression Techniques}} like quantization and pruning to optimize the pareto-optimal DL model configurations for embedded deployment on a given edgeAI platform, while minimizing resource consumption and latency.

    \item\underline{\textit{Full System Integration and Validation}} using a full-functional in-house fabricated prosthetic arm prototype of \textit{CognitiveArm}, while interfacing it with the OpenBCI UltraCortex Mark IV EEG headset. Our results demonstrate significant improvements in accuracy (up to 90\%), response time, and system performance, underscoring the practicality of the system for real-world prosthetic control on NVIDIA Jetson Orin Nano.
     \item\underline{\textit{Voice Command Integration}} for seamless mode switching, enabling control of the prosthetic arm's 3 DoF and facilitating multi-action control for various everyday tasks (e.g., handshake, cup picking) along with model validation.
    
\end{itemize}

\textbf{Paper Organization:} Section II discusses related work. Section III presents the CognitiveArm system in further detail. Section IV details the Exploration Search, while  Section V describes the experimental setup, followed by results and discussion in Section VI. We conclude in Section VII.

\section{Related Work}
Non-invasive brain-computer interfaces (BCIs) have become a focal point in prosthetic control due to their safety and accessibility compared to invasive methods. Invasive BCIs, like Neuralink's implanted electrodes, offer high-precision control but come with surgical risks and high costs \cite{Musk2019-hl}. Non-invasive systems, such as those developed by OpenBCI \cite{OPENBCI} and Emotiv \cite{Emotiv}, interpret EEG signals to control prosthetics and other devices \cite{veena2020review}. However, these systems often face challenges with signal quality, limited channel counts, and the need for extensive user-specific calibration. Table II summarizes the performance of CognitiveArm against existing EEG-based prosthetic systems. CognitiveArm achieves higher accuracy (90\%) and significantly lower latency due to model compression techniques (70\% pruning, quantization). Unlike cloud-dependent systems, our on-device processing ensures real-time operation while maintaining computational efficiency. 

\footnotetext[1]{\textbf{High Accuracy:} Classification accuracy exceeding 90\% for all classes across test participants. \textbf{Moderate Accuracy:} Accuracy ranging between 75–90\%, indicating improvable performance.}

\footnotetext[2]{\textbf{High Cost:} Products or services priced above \$5000. \textbf{Moderate Cost:} Prices ranging between \$1000–\$5000, representing a mid-range category. \textbf{Low Cost:} Prices below \$1000, signifying affordability.}


\begin{table}[h!]
\centering
\scriptsize
\setlength{\tabcolsep}{2pt} 
\caption{Comparison of Brain-Controlled Prosthetic Arms}
\begin{tabular}{|p{1.9cm}|p{1.5cm}|p{0.6cm}|p{0.82cm}|p{3.5cm}|}
\hline
\textbf{Solution} & \textbf{Method} & \textbf{Acc.\footnotemark[1]} & \textbf{Cost\footnotemark[2]} & \textbf{Scope} \\ 
\hline
\cite{Ali2020} & EEG-based & Mod. & Low & Limited real-time use \\ 
\hline
\cite{Chinbat2018} & EEG-based & Mod. & High & Limited real-time use \\ 
\hline
\cite{Beyrouthy2016} & EEG-based & Mod. & High & Power-intensive, limited use \\ 
\hline
\cite{Lonsdale2020} & EEG + sEMG & High & Mod. & High resource demand \\ 
\hline
\cite{Zhang2024} & EEG + EoG & 80\% & Mod. & Simple movements, user-dependent \\ 
\hline
\cite{Vilela2020} & EEG-based & High & High & Invasive Solution \\ 
\hline
\cite{MindArm} & EEG-based & 87.5\% & Low & Affordable, modular \\ 
\hline
\cite{NeuroLimb} & EEG + sEMG & High & Low & Designed for developing regions \\ 
\hline
BeBionic \cite{bionic} & sEMG-based & High & £30k & More Grips, fine motor control  \\ 
\hline
LUKE Arm \cite{luke-arm} & sEMG-based & High & \$50k+ & Powered joints, fine motor control\\ 
\hline
i-Limb \cite{i-limb} & sEMG-based & High & \$40-50k & Multi-articulating, customizable \\ 
\hline
Michelangelo \cite{Michelangelo} & sEMG-based & High & \$50k+ & Advanced control, multiple grips \\ 
\hline
Shadow Hand \cite{ShadowHand} & sEMG-based & High & \$65k+ & High dexterity, advanced robotics \\ 
\hline
\textbf{Congitive Arm} & EEG-based & High & \$500 & 3 DoF, efficient implementation \\ 
\hline
\end{tabular}
\end{table}


BCI technologies have shown potential in controlling prosthetic arms, providing mobility solutions for individuals with neurological or physical impairments \cite{Chen2023}. Studies explore both invasive and non-invasive approaches to brain-controlled prosthetics \cite{Kumari2023, He2024}. Several studies developed an EEG-based BCI system for prosthetic control aimed at ALS and spinal injury patients \cite{Ali2020, Chinbat2018}. However, they lack computationally efficient solution with simpler control methods described. Another research project on an EEG mind-controlled robotic arm \cite{Beyrouthy2016} demonstrated hand movements, but suffered from high power consumption and inefficiencies. Further advancements have incorporated DL and 3D printing, such as in a cost-effective 3D-printed robotic arm controlled by sEMG sensors and DL \cite{Abbady2022}. \textit{While promising, the model remains resource-hungry}. A separate work on brain-controlled prosthetic hands \cite{Zhang2024} demonstrated varied user-dependent accuracy (80\%) for simple finger movements.

Machine learning techniques have also been explored in prosthetics. A comparison of KNN, SVM, and LDA classifiers for BCI control found LDA achieving up to 87.5\% accuracy for movement classification \cite{make3040042}. A comprehensive review of BCIs in prosthetic control highlighted signal acquisition and real-time performance challenges. Innovative solutions like MindArm \cite{MindArm} and LIBRA NeuroLimb \cite{NeuroLimb} emphasize affordability and accessibility in prosthetics. However, our work distinguishes itself by integrating a real-time, EEG-based control system optimized for low-latency and high-efficiency DNN. While these systems offer a cost-effective DNN-based solution, it lacks comprehensive integration of control efficiency and timing, which our system addresses to ensure precise and rapid response in real-world prosthetic applications.

\section{CognitiveArm: Design and Methodology}
\label{sec:methodology}

\begin{figure}[ht]
    \centering
    \includegraphics[width=1\linewidth]{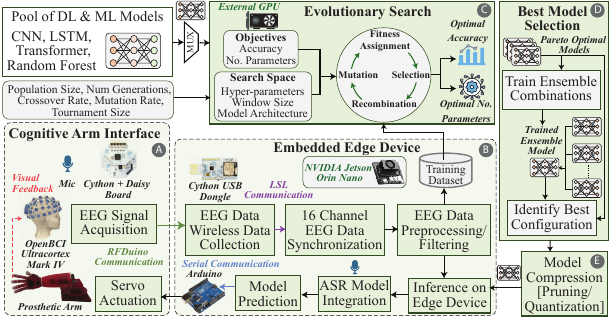}
    \caption{\textit{CognitiveArm} system methodology showcasing EEG signal acquisition, wireless data synchronization, preprocessing and filtering, evolutionary search for Pareto-optimal DL/ML models, ensemble training and best model selection, and real-time control of the prosthetic arm on an embedded edge device with integrated ASR-based voice commands.   
    \label{fig:methodology}}   
\end{figure}

The methodology for the \textit{CognitiveArm} system encompasses the entire pipeline from EEG data acquisition to real-time prosthetic arm control. The key components include EEG signal acquisition and synchronization, data preprocessing and filtering, EEG dataset generation and annotation, model selection and optimization through evolutionary search, model training and evaluation, and integration with automatic speech recognition for enhanced control. The methodology of the CongnitiveArm is illustrated in Figure~\ref{fig:methodology}.

\subsection{EEG Data Acquisition and Preprocessing}
\label{subsec:data_acquisition}

\subsubsection{EEG Data Acquisition}

\begin{figure}[ht]
    \centering
    \includegraphics[width=1\linewidth]{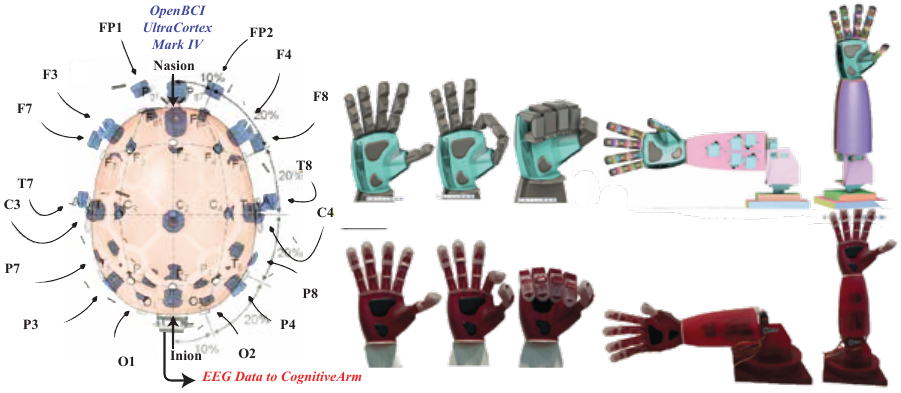}
    \caption{EEG data is collected using the 10-20 electrode system, incorporating 16 electrodes for comprehensive brainwave acquisition. The recorded EEG signals are transmitted to the \textit{CognitiveArm} for processing, where the system interprets the brain signals to control and demonstrate various hand poses.}
    \label{fig:10-20}
\end{figure}

EEG data were acquired using the OpenBCI Ultracortex Mark IV EEG headset paired with 16-channel Cyton + Daisy Biosensing boards, which stream EEG data from 16 electrodes placed according to the 10-20 system (shown in Figure~\ref{fig:10-20}). This setup captures neural activity from key motor-related regions such as FP1, FP2, C3, and C4. The use of dry electrodes and multiple channels enables comprehensive, non-intrusive brain signal collection, ideal for DL feature extraction. Data acquisition was facilitated using BrainFlow~\cite{brainflow}, an open-source library optimized for high-fidelity EEG acquisition and streaming. BrainFlow's board-agnostic design and multi-threaded data streaming capabilities provide enhanced flexibility and efficiency, crucial for real-time control.

\begin{figure}[ht]
    \centering
    \includegraphics[width=1\linewidth]{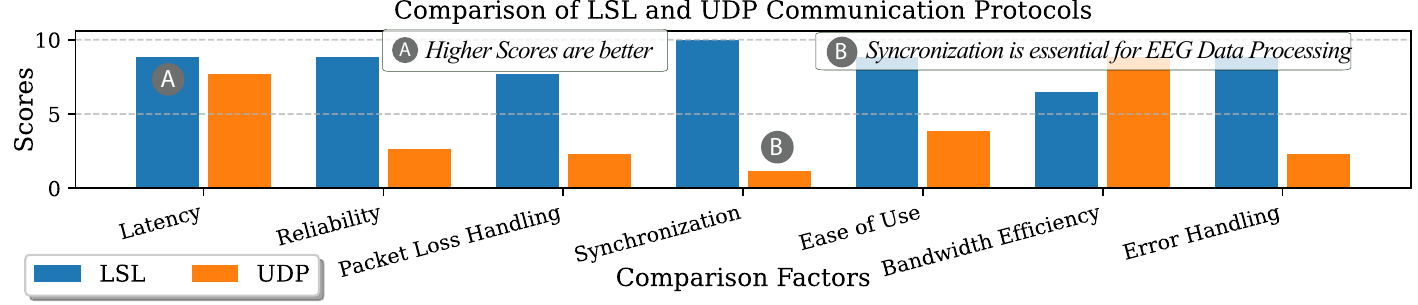}
    \caption{Comparison of LSL and UDP, showing LSL's strengths in key factors, while UDP excels in bandwidth efficiency only.}
    \label{fig:LSL_UDP}    
\end{figure}

\subsubsection{Signal Acquisition and Synchronization}

The EEG signals were collected and streamed using the Lab Streaming Layer (LSL) communication protocol~\cite{Kothe2024}, chosen for its low latency and high sample rate, which are critical for capturing rapid changes in brainwave patterns. LSL allows consistent data streaming at 125\,Hz, ensuring precise synchronization and time-stamping, essential for accurate EEG signal acquisition in real-time setups. Figure~\ref{fig:LSL_UDP} compares LSL with UDP, highlighting LSL's advantages in several key areas.

\subsubsection{Preprocessing and Filtering}

\begin{figure}[ht]
    \centering
    \includegraphics[width=1\linewidth]{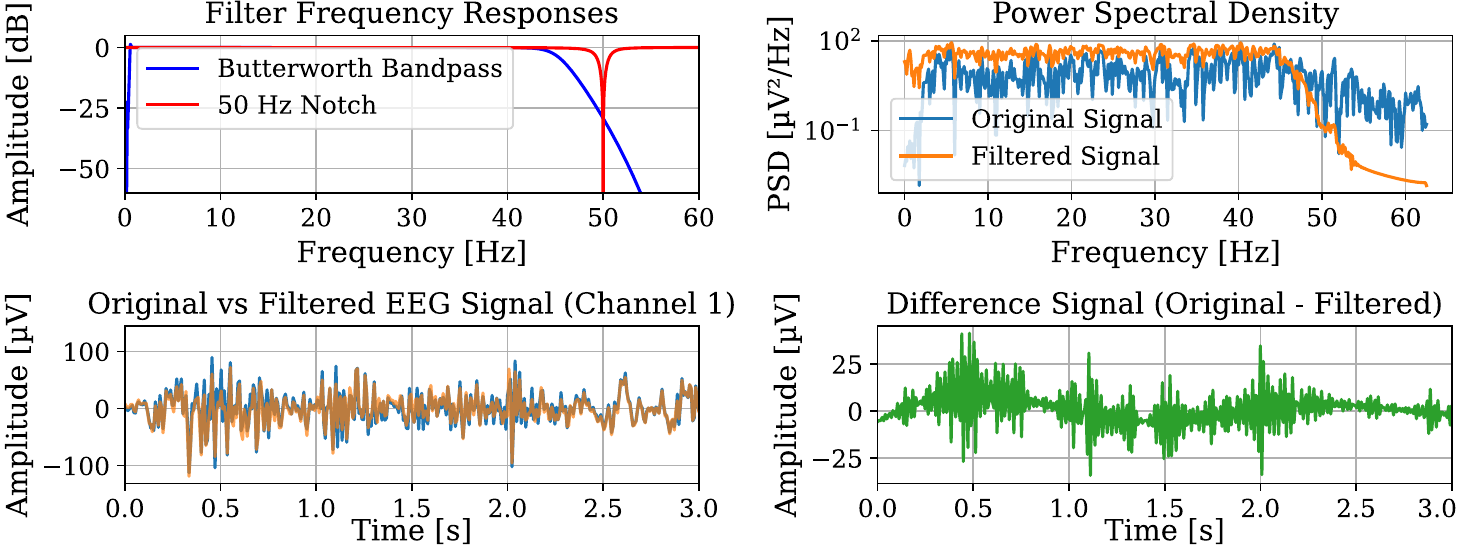}
    \caption{Comparison of original and filtered EEG signals using a Butterworth bandpass filter and notch filter, demonstrating the removal of noise and retention of relevant brainwave frequencies for a single channel.}
    \label{fig:Filter}    
\end{figure}

To ensure high-quality EEG data for reliable prosthetic control, a series of preprocessing steps were implemented aimed at removing noise and artifacts:

\begin{itemize}
    \item \textbf{Butterworth Bandpass Filter}: A 9th-order Butterworth bandpass filter was applied to retain the critical frequency range of 0.5--45\,Hz, encompassing key brainwave activities (delta, theta, alpha, beta waves). This filtering process effectively removes low-frequency drift and high-frequency noise.
    
    \item \textbf{Notch Filter}: A 50\,Hz notch filter with a quality factor of 30 was used to eliminate powerline interference, removing electrical noise from power sources that could contaminate the EEG signals.
    
    \item \textbf{Artifact Removal}: Standard signal cleaning techniques provided by BrainFlow were employed to address common EEG artifacts such as eye blinks and muscle movements, enhancing the signal-to-noise ratio (SNR) for accurate classification by the deep learning models.
The result of filtering is illustrated in Figure~\ref{fig:Filter}.
\end{itemize}

\begin{figure*}[ht]
    \centering
    \includegraphics[width=1\linewidth]{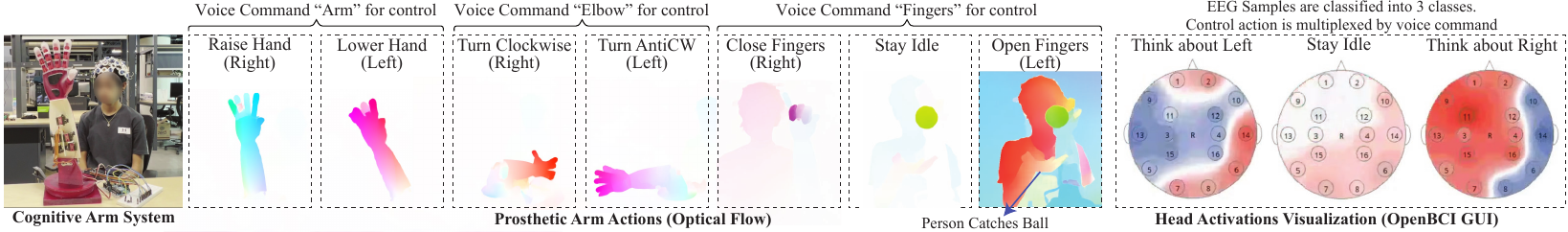}
    \caption{Movement of the arm in correspondence the user's EEG signals is indicated. The optical flow images are generated using FlowFormer++ model \cite{shi2023flowformermaskedcostvolume}.}
    \label{fig:brainmap}
\end{figure*}

\subsection{EEG Dataset Generation and Annotation}
\label{subsec:data_generation}

\subsubsection{Experimental Protocol}

Experiments involved five participants (four males and one female), aged between 18 and 31, all healthy individuals with no known neurological conditions. Participants were familiarized with the experimental setup and instructed to perform \textbf{mental tasks} corresponding to \textit{three classes of actions} detailed below:

\begin{itemize}
        \item \textit{Move Right}: Participants imagined moving their right hand to the right without physical movement.
        \item \textit{Move Left}: Participants imagined moving their left hand to the left without physical movement.
        \item \textit{Stay Idle}: Participants maintained a calm state of mind without focusing on any particular movement.
    \end{itemize}
\textbf{Collection Structure}: Each mental task was performed for 10 seconds, followed by a 10-second idle period, forming a structured sequence. This process was repeated to collect approximately 5 minutes of EEG data per participant across three separate sessions.



\subsubsection{Data Labeling}

EEG data were labeled based on auditory cues (beep sounds) indicating the start of each mental task. Labels were assigned for each mental task session as having one label, which was later split into sliding windows each having the same label as the mental task. Potential delays between the auditory cue and participant response were accounted for by including transition periods in the labeled data. Participant feedback was obtained to confirm their mental focus during data collection, ensuring label accuracy.

\subsubsection{Data Segmentation}

The preprocessed EEG data were segmented into sliding windows to capture temporal dynamics:

\begin{itemize}
    \item \textbf{Window Size}: Varying sizes ranging from 100 to 200 samples (0.8\,s to 1.6\,s at 125\,Hz) were tested.
    \item \textbf{Sliding Step}: A step size of 25 samples (0.2\,s) was used.
\end{itemize}

This resulted in a dataset suitable for input into the DL/ML models.

\subsection{Model Selection and Optimization}
\label{subsec:model_selection}

\subsubsection{Embodied DL/ML Model Selection}

A set of DL/ML architectures was evaluated for EEG signal classification, including:

\begin{itemize}
    \item Convolutional Neural Networks (CNNs) / Part of Ensemble
    \item Long Short-Term Memory (LSTM) networks / Part of Ensemble
    \item Transformers / Part of Ensemble
    \item Random Forest classifiers / Part of Ensemble
\end{itemize}

The goal was to balance classification accuracy with computational efficiency for real-time deployment on edge devices.

\subsubsection{Evolutionary Search Algorithm}

An evolutionary search algorithm was employed to identify Pareto-optimal model configurations, balancing performance and efficiency.

\paragraph{Problem Formulation}

Let $M$ represent the set of candidate models, each defined by architecture and hyperparameters. Models were optimized for two objectives: maximizing accuracy $A(m_i)$ and minimizing parameter count $P(m_i)$. The aim is to find models that offer the best trade-off between these objectives.

\paragraph{Algorithm Steps}
We employ an evolutionary algorithm that operates on a population of models, evolving them over generations. The process is detailed below:

\begin{itemize}
    \item \textbf{Initialization}: A population $P_0$ of $N$ models with varied architectures and hyperparameters was generated randomly.
    \item \textbf{Fitness Evaluation}: Each model's fitness $m_i$ was determined based on validation accuracy and parameter count, using a scoring function that balances these objectives with weights $w_A$, $w_P$:
        
    \begingroup
    \vspace{-5pt}
    \setlength{\abovedisplayskip}{10pt} 
    \setlength{\belowdisplayskip}{10pt} 
    \scriptsize
    \[
    S(m_i) = w_A \cdot \frac{A(m_i) - \min(A)}{\max(A) - \min(A)} - w_P \cdot \frac{P(m_i) - \min(P)}{\max(P) - \min(P)}
    \]
    \endgroup
    \item \textbf{Selection}: Tournament selection was employed to choose parent models for reproduction based on fitness score $S(m_i)$.
    \item \textbf{Crossover and Mutation}: Evolutionary operations were applied to evolve the population over $G$ generations, introducing diversity and preventing premature convergence.
    \item \textbf{Pareto Front Identification}: After $G$ generations, models offering the best trade-off between accuracy and efficiency were identified. Pareto Front is evaluated using the following criteria:
    
   
    \begingroup
    \vspace{-10pt}
    
    \setlength{\abovedisplayskip}{10pt} 
    \setlength{\belowdisplayskip}{10pt} 
    \scriptsize
    \[
    \hspace*{-4mm} 
    {\footnotesize F = \{m_i \in P_G \mid \not\exists m_j \in P_G \text{ : } A(m_j) > A(m_i) \text{ \& } P(m_j) \leq P(m_i)\}}
    \]
    \endgroup
    \item \textbf{Best Model Selection}: Models meeting or exceeding an accuracy threshold $\alpha$ were selected based on minimal parameter count.        
    
    \begingroup
    \vspace{-10pt}
    \setlength{\abovedisplayskip}{10pt} 
    \setlength{\belowdisplayskip}{10pt} 
    \scriptsize
    \[
    m_{\text{best}} = 
    \begin{cases}
    \arg\min_{m_i \in F, A(m_i) \geq \alpha} P(m_i), & \text{if accuracy constraint met}\\
    \arg\max_{m_i \in F} A(m_i), & \text{otherwise}
    \end{cases}
    \]
    \endgroup
\end{itemize}

Algorithm~1 details the evolutionary search for model selection.

\begin{algorithm}[h!]
\scriptsize
\caption{\small Evolutionary Search for Pareto-Optimal Model}
\begin{algorithmic}[1]
\State \textbf{Input:} Population size $N$, generations $G$, accuracy threshold $\alpha$, mutation rate $p_m$, crossover rate $p_c$
\State \textbf{Output:} Pareto-optimal model $m_{\text{best}}$
\State Initialize population $P_0$ of size $N$
\For{generation $g = 1$ to $G$}
    \For{each model $m_i \in P_g$}
        \State Train and evaluate $m_i$ to compute $A(m_i)$ and $P(m_i)$
        \State Compute fitness $S(m_i)$ based on accuracy and parameter count
    \EndFor
    \State Select parents using tournament selection
    \State Apply crossover with probability $p_c$
    \State Apply mutation with probability $p_m$
    \State Update population $P_{g+1}$ with offspring
\EndFor
\State Compute Pareto front $F$ from $P_G$
\If{$\exists m_i \in F$ such that $A(m_i) \geq \alpha$}
    \State $m_{\text{best}} = \arg\min_{m_i \in F, A(m_i) \geq \alpha} P(m_i)$
\Else
    \State $m_{\text{best}} = \arg\max_{m_i \in F} A(m_i)$
\EndIf
\State \Return $m_{\text{best}}$
\end{algorithmic}
\end{algorithm}

\paragraph{Hyperparameter Tuning}

Hyperparameters such as window size, learning rate, number of layers, hidden units, and dropout rates were optimized. Table~\ref{tab:hyperparameters} summarizes the hyperparameters and model architectures tested during the evolutionary search.

\begin{table*}[h!]
\centering
\footnotesize
\setlength{\tabcolsep}{2pt} 
\caption{Hyperparameters and Model Architectures Tested in Evolutionary Search}
\label{tab:hyperparameters}
\begin{tabular}{|c|c|c|c|c|}
\hline
\textbf{Model} & \textbf{Architecture} & \textbf{Hyperparameters Tested} & \textbf{Optimizers} & \textbf{Other Details} \\ \hline
\textbf{LSTM} & 64, 128, 256 Units & Window Size (100--200), Dropout (0.1--0.5), Hidden Layers (1--3) & Adam, RMSProp & Learning Rate (1e-3--1e-5) \\ \hline
\textbf{CNN} & 2--4 Conv Layers & Filter Sizes (3x3, 5x5), Pooling (Max/Avg), Stride (1--2) & Adam, SGD & Batch Size (32--128) \\ \hline
\textbf{Random Forest} & 100--500 Trees & Features (Mean, Std, Min, Max, Var), Max Depth (10--None) & N/A (Non-Gradient) & Feature Selection \\ \hline
\textbf{Transformer} & 2--6 Layers & Attention Heads (2--8), Hidden Units (64--256), Dropout (0.1--0.5) & AdamW & Weight Decay (1e-4--1e-6) \\ \hline
\end{tabular}
\end{table*}

\subsection{Model Training and Evaluation}
\label{subsec:model_training}

\subsubsection{Cross-Subject Validation}


To evaluate model generalization, a leave-one-subject-out cross-validation strategy was employed. Data from four participants were used for training and validation (80:20 split), while the fifth participant's data was reserved for testing on unseen samples. Real-time validation was conducted with the prosthetic arm, where the DL models predicted control actions and translated them into corresponding motions. Ground truth labels for mental tasks were assigned based on vocal cues provided by participants during the session, ensuring accuracy in action validation. This process was repeated with each participant serving as the test subject, providing comprehensive validation across individuals.

\subsubsection{Evaluation Metrics}

The primary metric for model performance was classification accuracy, computed as the percentage of correctly classified windows over the total number of windows in the test set. To assess the statistical significance of the results, we calculated the mean accuracy and standard deviation across different test subjects.

\subsubsection{Overfitting Analysis}

To mitigate overfitting, we monitored training and validation losses to detect divergence and employed a leave-one-subject-out cross-validation strategy to ensure generalization to unseen participants. Regularization techniques, a balanced dataset, and a sliding window approach minimized bias across classes and participants. Additionally, the ensemble DL model enhanced robustness and overall performance by aggregating predictions and reducing individual variances.



\subsubsection{Ensuring Label Accuracy}
To ensure label accuracy, several protocols were implemented. Transition periods were carefully incorporated to account for potential delays between auditory cues and participant responses, mitigating the effects of auditory lag. Before data collection, participants underwent a guided tutorial to familiarize themselves with the setup and the task requirements. The dataset was balanced across the three classes—\textit{right}, \textit{left}, and \textit{idle}—to prevent model bias towards any particular action. Additionally, participants provided real-time feedback to confirm their mental focus during data collection, ensuring the reliability of the labels. These measures collectively enhanced the quality and consistency of the dataset.



\subsection{Optimizing Neural Network Architectures}
\label{subsec:nn_optimization}

After identifying the best-performing models, model compression techniques were applied to optimize them for embedded deployment:

\subsubsection{Pruning}

Neural network connections were pruned at different levels (90\%, 70\%, 50\%, 30\%, and 0\%), reducing the number of parameters without significant accuracy loss. Global pruning was applied to achieve a uniform reduction across the network.

\subsubsection{Quantization}

Post-training quantization converted model weights into lower-precision formats (e.g., 8-bit integers), decreasing memory usage and inference time. This step further optimized the models for deployment on resource-constrained edge devices.

\subsection{Automatic Speech Recognition Integration}
\label{subsec:asr_integration}

An Automatic Speech Recognition (ASR) system using the Whisper-small model \cite{radford2022} was integrated to enhance user interaction:

\subsubsection{Voice Commands}

The ASR system enabled voice commands for seamless mode switching, allowing users to control the prosthetic arm's degrees of freedom (e.g., ``elbow,'' ``arm,'' ``fingers''). This facilitated multi-action control for various everyday tasks.
\begin{figure}[ht]
    \centering
    \includegraphics[width=1\linewidth]{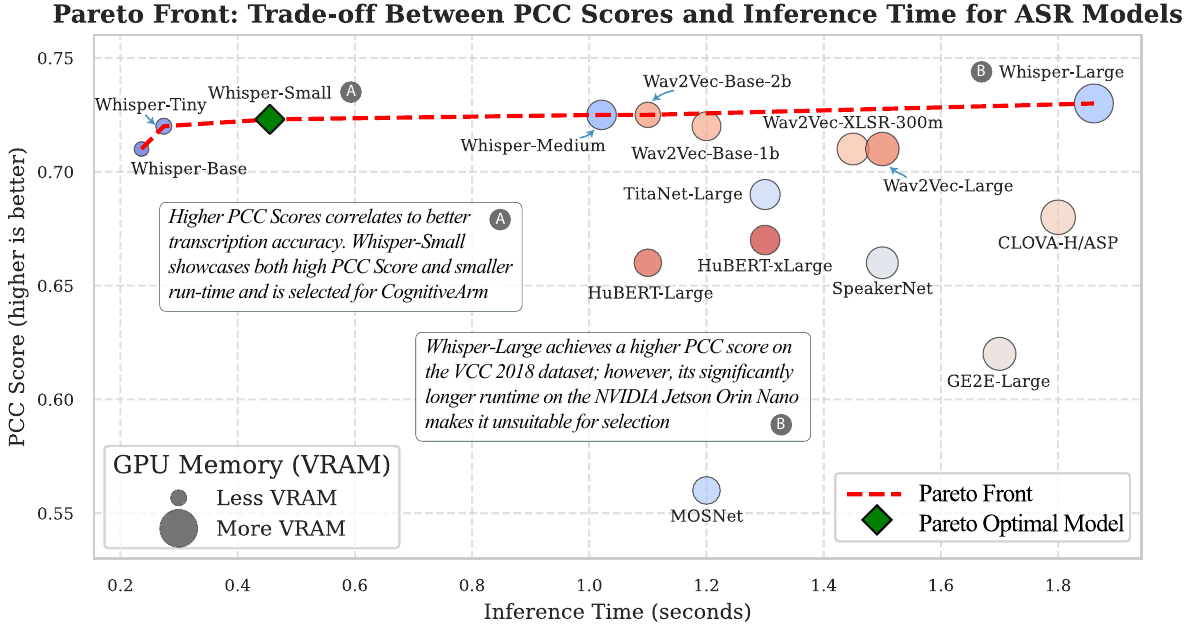}
    \caption{Pareto front illustrating the trade-off between PCC scores and inference time for various ASR models. The size of each marker represents the VRAM usage (GPU memory) for each model. The Pareto front (red line) highlights the models offering the best balance between performance and resource efficiency, with the optimal model shown in green. }
    \label{fig:whisper}    
\end{figure}
\subsubsection{Voice Activity Detection}

A Voice Activity Detection (VAD) algorithm was employed to trigger the ASR model only when speech was detected, minimizing resource consumption and latency.

\subsubsection{Parallel Processing}

The ASR system ran in a separate thread from the EEG classification, maintaining real-time responsiveness and ensuring that the EEG processing was not interrupted.

\subsubsection{System Integration and Real-Time Control}
The optimized models and ASR system were deployed on edge AI hardware, enabling the \textit{CognitiveArm} system to operate independently with low latency, facilitating real-time interaction and control of the prosthetic arm. The end-to-end system integrates sensing, decision-making, motor control, and actuation, ensuring seamless operation in real-world scenarios.

\section{Experimental Setup}
The experimental setup for cognitive Arm integrates several key components, including model training and deployment, signal processing, and hardware control mechanisms. The system is designed for real-time EEG-based control of prosthetic movements, enhanced by speech recognition capabilities. This section details the hardware, software, and models used in the experimental setup.
\subsection{Prosthetic Arm Design and Prototyping}
The prosthetic arm was designed using Autodesk Fusion 360, with three DoF: elbow flexion/extension and finger movements for gripping. The design was iteratively optimized for mechanical efficiency, durability, and functionality. Components were 3D printed using the Stratasys J750, ensuring precision and accurate material properties. The assembled arm features five embedded servo motors controlling finger movements for precise actions like gripping shown in Figure~\ref{fig:10-20}.

\begin{figure*}[t]
    \centering
    \includegraphics[width=1\linewidth]{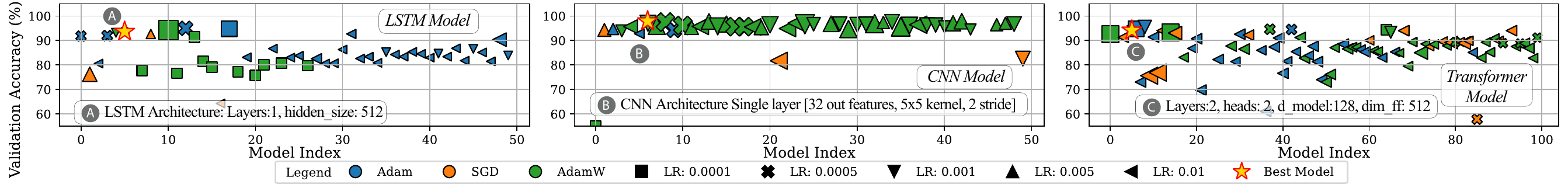}
    \caption{Results highlight the selection of hyperparameters and model architecutes using evolutionary search, balancing accuracy and computational efficiency, and provides insights into the performance of CNN, LSTM, and Transformer models. The points highlighted are the pareto optimal models individually.}
    \label{fig:idx_transfprmer}    
\end{figure*}
\begin{figure*}[t]
    \centering
    \includegraphics[width=1\linewidth]{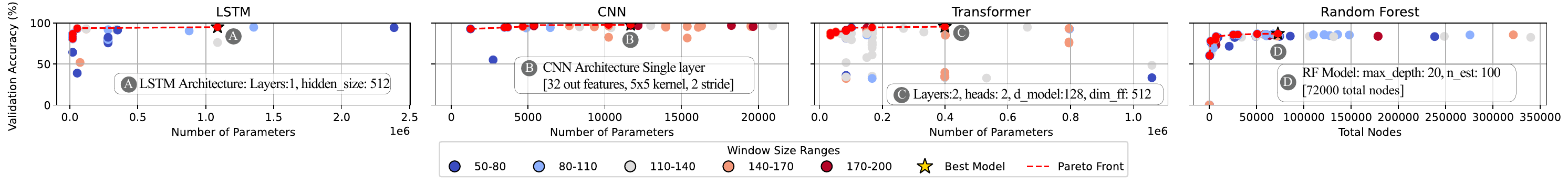}
    \caption{Pareto front showcasing the trade-off between accuracy and parameter size for different model configurations. Models on the front achieve optimal balance, demonstrating the best accuracy given their parameter count. Particularly CNN models achieve high validation accuracy.}
    \label{fig:idx_transfprmer}    
\end{figure*}
\subsubsection{EEG Signal Acquisition and Processing} EEG signals are collected using the Ultracortex Mark IV headset with Cyton + Daisy boards, streaming data using BrainFlow for real-time inference. Preprocessing involves Butterworth bandpass filtering and a 50 Hz notch filter to remove noise and artifacts.

\subsubsection{Model Training and Deployment} Neural network models (LSTM, CNN, Random Forest, and Transformer) are trained on the RTX A6000 Ada GPU to handle large training datasets efficiently. For edge deployment, the \textbf{NVIDIA Jetson Orin Nano} is used for real-time inference and prosthetic control, providing a balance of computational power and energy efficiency.

\subsubsection{Generating Action Labels} Trained models classify EEG signals into action labels (e.g., move left, move right, stay idle) at 15 Hz. These labels are translated into motor control commands for the prosthetic arm.

\subsubsection{Prosthetic Arm Control} Action labels are transmitted to an Arduino microcontroller, which sends precise motor control signals to servo motors, executing commands like gripping or elbow movement.

\subsubsection{Real-World Validation}
To validate the system in real-world scenarios, the EEG classification output was directly interfaced with the prosthetic arm. Participants independently controlled the arm's movements during test sessions, successfully translating their intended actions in 19 out of 20 sessions. For added reliability, participants verbally confirmed their intentions (e.g., saying \textit{right} for a right-hand action), which was synchronized with EEG labels. This protocol minimized ambiguity and ensured consistent evaluation.

\subsubsection{Servo Calibration} Servo motors are calibrated with a CCPM 3-channel tester to ensure alignment and consistent movement, optimizing the range of motion.

\subsubsection{System Communication and Integration}
The communication between the Jetson Orin Nano and the Arduino microcontroller is established via a serial communication protocol. The neural network models deployed on the Jetson Orin Nano process EEG signals and generate action labels in real-time, which are then transmitted to the Arduino to control the prosthetic’s motors, performing classification, and executing control commands.

\subsubsection{Safety and Comfort Measures}
Safety protocols were followed to ensure: Proper handling of the EEG headset and electrodes. Safe operation of the prosthetic arm, avoiding rapid or unexpected movements. Regular breaks between sessions to prevent fatigue and maintain concentration levels.

\begin{figure}[t]
    \centering
    \includegraphics[width=1\linewidth]{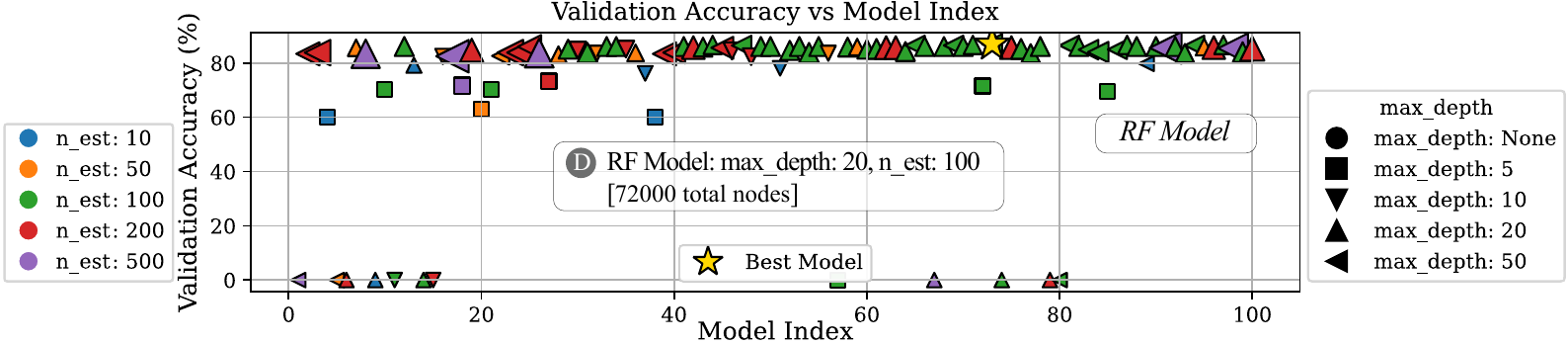}
    \caption{Selection of hyperparameters using evolutionary search, balancing number of estimators and tree depth for Random Forest.}
    \label{fig:idx_transfprmer}    
\end{figure}

\begin{figure}[t]
    \centering
    \includegraphics[width=0.9\linewidth]{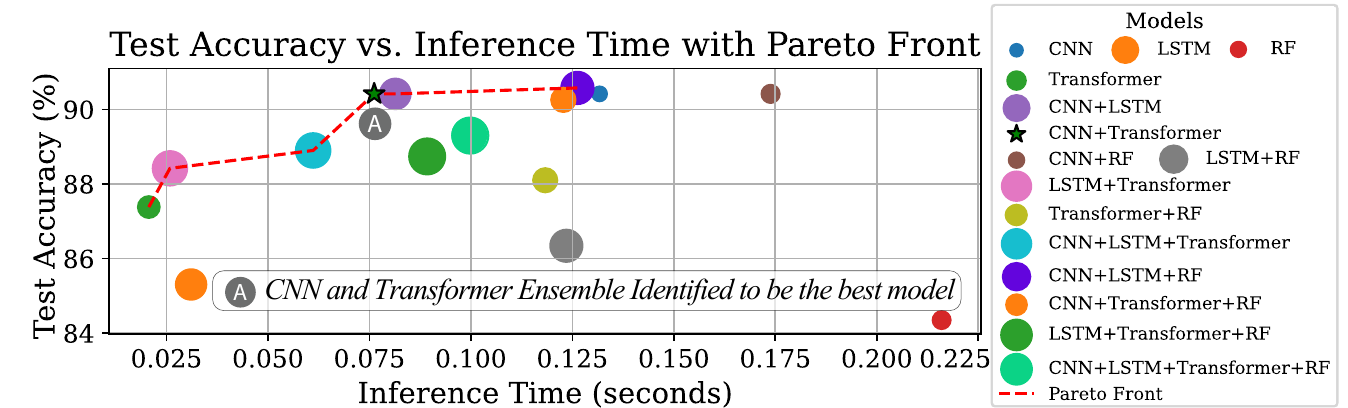}
    \caption{comparing ensemble models based on their inference time and accuracy, showcasing the best-performing ensemble of CNN and Transformer models. The highlighted ensemble demonstrates the most optimal balance between quick response times and high accuracy, indicating its suitability for real-time applications.}
    \label{fig:idx_transfprmer}    
\end{figure}

\begin{figure}[t]
    \centering
    \includegraphics[width=1\linewidth]{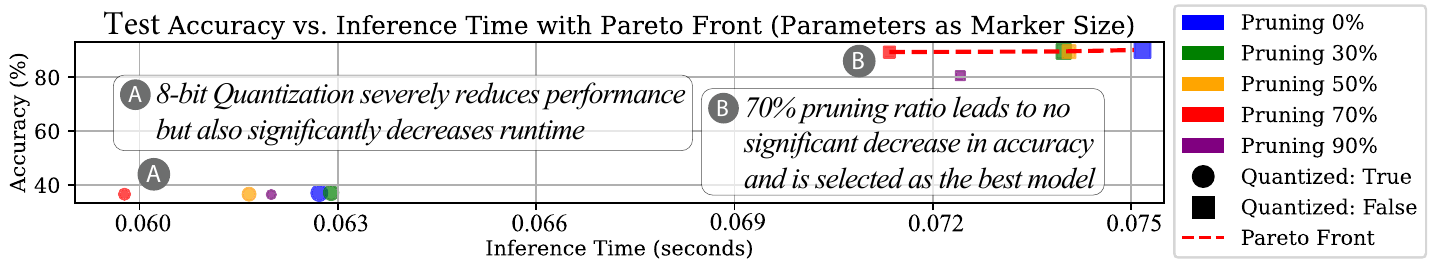}
    \caption{ Test Accuracy vs. Inference Time with Pareto Front. The 70\% pruned model (B) maintains high accuracy (above 90\%) with reduced runtime, while 8-bit quantization (A) significantly decreases runtime at the cost of accuracy. Marker size indicates model parameters.}
    \label{fig:idx_transfprmer}    
\end{figure}

\section{Results and Analysis}
\subsection{Statistical Analysis}
To evaluate cross-subject variability, we normalized EEG data using the mean and standard deviation of each participant’s readings. Statistical significance was assessed using paired t-tests to compare model performances across subjects. Additionally, confidence intervals (91\%) were computed for the test accuracies of each model. For ensemble models, variance reduction was analyzed to confirm the robustness of predictions against user-specific noise.

We identified the best-performing models for EEG signal classification: a CNN with one convolutional layer (32 filters, kernel size 5, stride 2) using a window size of 190; an LSTM with a single layer of 512 hidden units and a window size of 130; a Transformer with 2 layers, 2 heads, a model dimension of 128, and a window size of 190; and a Random Forest with 200 estimators and a window size of 90. 
The results highlight the effectiveness of combining model architectures and compression techniques for EEG-driven prosthetic control. The CNN and Transformer ensemble achieved the best trade-off, with 91\% accuracy and 0.075s inference time, ideal for real-time systems. Pruning further reduced inference time to 0.071 s while maintaining 90.1\% accuracy, demonstrating its practicality for edge devices. Quantization improved latency to 0.036 s with significant accuracy reduction (38.5\%), unsuitable for safety-critical applications. These findings validate the CognitiveArm system’s efficiency and adaptability for resource-constrained environments.


\section{Conclusion}

This paper presented an intelligent prosthetic arm system that integrates advanced EEG signal processing with embedded deep learning (DL) models, specifically optimized for real-time control and resource-constrained environments. Our approach leverages model compression techniques, including pruning and quantization, to significantly reduce latency and computational overhead while maintaining high accuracy. Through an evolutionary search method, we identified Pareto-optimal DL configurations that strike a balance between computational efficiency and predictive performance. The 70\% pruned model demonstrated exceptional performance, achieving 90.1\% accuracy while reducing inference time to enable real-time operation. The combination of architecture optimization, model compression, and robust evaluation across diverse participants ensures the system’s adaptability to real-world applications. Furthermore, the proposed framework highlights the potential for scalable brain-computer interface (BCI) systems capable of delivering reliable, low-latency control for prosthetic devices. This work paves the way for future advancements in EEG-driven prosthetics, emphasizing efficiency, scalability, and practical deployment in everyday scenarios.

\end{spacing}
\section*{Acknowledgment}
This work was partially supported by the NYUAD Center for Artificial Intelligence and Robotics (CAIR), funded by Tamkeen under the NYUAD Research Institute Award CG010.

\begingroup
\scriptsize
\bibliographystyle{IEEEtran}
\bibliography{cite}

\begin{thebibliography}{10}
\providecommand{\url}[1]{#1}
\csname url@samestyle\endcsname
\providecommand{\newblock}{\relax}
\providecommand{\bibinfo}[2]{#2}
\providecommand{\BIBentrySTDinterwordspacing}{\spaceskip=0pt\relax}
\providecommand{\BIBentryALTinterwordstretchfactor}{4}
\providecommand{\BIBentryALTinterwordspacing}{\spaceskip=\fontdimen2\font plus
\BIBentryALTinterwordstretchfactor\fontdimen3\font minus \fontdimen4\font\relax}
\providecommand{\BIBforeignlanguage}[2]{{%
\expandafter\ifx\csname l@#1\endcsname\relax
\typeout{** WARNING: IEEEtran.bst: No hyphenation pattern has been}%
\typeout{** loaded for the language `#1'. Using the pattern for}%
\typeout{** the default language instead.}%
\else
\language=\csname l@#1\endcsname
\fi
#2}}
\providecommand{\BIBdecl}{\relax}
\BIBdecl

\bibitem{fitzgibbons2015functional}
P.~Fitzgibbons and G.~Medvedev, ``Functional and clinical outcomes of upper extremity amputation,'' \emph{JAAOS-Journal of the American Academy of Orthopaedic Surgeons}, vol.~23, no.~12, pp. 751--760, 2015.

\bibitem{Padfield2022}
\BIBentryALTinterwordspacing
N.~Padfield, K.~Camilleri, T.~Camilleri, S.~Fabri, and M.~Bugeja, ``A comprehensive review of endogenous eeg-based bcis for dynamic device control,'' \emph{Sensors 2022, Vol. 22, Page 5802}, vol.~22, p. 5802, 8 2022. [Online]. Available: \url{https://www.mdpi.com/1424-8220/22/15/5802/htm https://www.mdpi.com/1424-8220/22/15/5802}
\BIBentrySTDinterwordspacing

\bibitem{Wang2019}
Y.~Wang, M.~Nakanishi, and D.~Zhang, ``Eeg-based brain-computer interfaces,'' \emph{Advances in Experimental Medicine and Biology}, vol. 1101, pp. 41--65, 2019.

\bibitem{Ma2024}
\BIBentryALTinterwordspacing
Z.~Z. Ma, J.~J. Wu, Z.~Cao, X.~Y. Hua, M.~X. Zheng, X.~X. Xing, J.~Ma, and J.~G. Xu, ``Motor imagery-based brain–computer interface rehabilitation programs enhance upper extremity performance and cortical activation in stroke patients,'' \emph{Journal of NeuroEngineering and Rehabilitation}, vol.~21, pp. 1--14, 12 2024. [Online]. Available: \url{https://jneuroengrehab.biomedcentral.com/articles/10.1186/s12984-024-01387-w}
\BIBentrySTDinterwordspacing

\bibitem{trial}
\BIBentryALTinterwordspacing
``Trial: Non-invasive bci-controlled assistive devices | als tdi.'' [Online]. Available: \url{https://www.als.net/als-trial-navigator/500010/}
\BIBentrySTDinterwordspacing

\bibitem{Leng2024}
\BIBentryALTinterwordspacing
J.~Leng, X.~Yu, C.~Wang, J.~Zhao, J.~Zhu, X.~Chen, Z.~Zhu, X.~Jiang, J.~Zhao, C.~Feng, Q.~Yang, J.~Li, L.~Jiang, F.~Xu, and Y.~Zhang, ``Functional connectivity of eeg motor rhythms after spinal cord injury,'' \emph{Cognitive Neurodynamics}, pp. 1--15, 6 2024. [Online]. Available: \url{https://link.springer.com/article/10.1007/s11571-024-10136-7}
\BIBentrySTDinterwordspacing

\bibitem{Li2023}
H.~Li, H.~Ji, J.~Yu, J.~Li, L.~Jin, L.~Liu, Z.~Bai, and C.~Ye, ``A sequential learning model with gnn for eeg-emg-based stroke rehabilitation bci,'' \emph{Frontiers in Neuroscience}, vol.~17, p. 1125230, 4 2023.

\bibitem{Maslova2023}
O.~Maslova, Y.~Komarova, N.~Shusharina, A.~Kolsanov, A.~Zakharov, E.~Garina, and V.~Pyatin, ``Non-invasive eeg-based bci spellers from the beginning to today: a mini-review,'' \emph{Frontiers in Human Neuroscience}, vol.~17, p. 1216648, 8 2023.

\bibitem{s21134293}
\BIBentryALTinterwordspacing
K.~Belwafi, S.~Gannouni, and H.~Aboalsamh, ``Embedded brain computer interface: State-of-the-art in research,'' \emph{Sensors}, vol.~21, no.~13, 2021. [Online]. Available: \url{https://www.mdpi.com/1424-8220/21/13/4293}
\BIBentrySTDinterwordspacing

\bibitem{app14146347}
W.~H. Elashmawi, A.~Ayman, M.~Antoun, H.~Mohamed, S.~E. Mohamed, H.~Amr, Y.~Talaat, and A.~Ali, ``A comprehensive review on brain–computer interface (bci)-based machine and deep learning algorithms for stroke rehabilitation,'' \emph{Applied Sciences}, vol.~14, no.~14, 2024.

\bibitem{Samal}
P.~Samal and M.~F. Hashmi, ``Role of machine learning and deep learning techniques in eeg‑based bci emotion recognition system: a review,'' \emph{Artificial Intelligence Review}, 02 2024.

\bibitem{Craik2019-mi}
A.~Craik, Y.~He, and J.~L. Contreras-Vidal, ``Deep learning for electroencephalogram ({EEG}) classification tasks: a review,'' \emph{Journal of Neural Engineering}, vol.~16, no.~3, p. 031001, Feb. 2019.

\bibitem{Roy_2019}
\BIBentryALTinterwordspacing
Y.~Roy, H.~Banville, I.~Albuquerque, A.~Gramfort, T.~H. Falk, and J.~Faubert, ``Deep learning-based electroencephalography analysis: a systematic review,'' \emph{Journal of Neural Engineering}, vol.~16, no.~5, p. 051001, aug 2019. [Online]. Available: \url{https://dx.doi.org/10.1088/1741-2552/ab260c}
\BIBentrySTDinterwordspacing

\bibitem{7994570}
N.~D. Lane, S.~Bhattacharya, A.~Mathur, P.~Georgiev, C.~Forlivesi, and F.~Kawsar, ``Squeezing deep learning into mobile and embedded devices,'' \emph{IEEE Pervasive Computing}, vol.~16, no.~3, pp. 82--88, 2017.

\bibitem{pandemic}
\BIBentryALTinterwordspacing
A.~Rodríguez, M.~Z. Islam, A.~S. M.~S. Sagar, and H.~S. Kim, ``Enabling pandemic-resilient healthcare: Edge-computing-assisted real-time elderly caring monitoring system,'' \emph{Applied Sciences 2024, Vol. 14, Page 8486}, vol.~14, p. 8486, 9 2024. [Online]. Available: \url{https://www.mdpi.com/2076-3417/14/18/8486/htm https://www.mdpi.com/2076-3417/14/18/8486}
\BIBentrySTDinterwordspacing

\bibitem{9892902}
M.~A. Hanif, G.~M. Sarda, A.~Marchisio, G.~Masera, M.~Martina, and M.~Shafique, ``Conlocnn: Exploiting correlation and non-uniform quantization for energy-efficient low-precision deep convolutional neural networks,'' in \emph{2022 International Joint Conference on Neural Networks (IJCNN)}, 2022, pp. 1--8.

\bibitem{9211496}
H.~Ahmad, T.~Arif, M.~A. Hanif, R.~Hafiz, and M.~Shafique, ``Superslash: A unified design space exploration and model compression methodology for design of deep learning accelerators with reduced off-chip memory access volume,'' \emph{IEEE Transactions on Computer-Aided Design of Integrated Circuits and Systems}, vol.~39, no.~11, pp. 4191--4204, 2020.

\bibitem{Musk2019-hl}
E.~Musk and {Neuralink}, ``\BIBforeignlanguage{en}{An integrated {Brain-Machine} interface platform with thousands of channels},'' \emph{\BIBforeignlanguage{en}{J Med Internet Res}}, vol.~21, no.~10, p. e16194, Oct. 2019.

\bibitem{OPENBCI}
\BIBentryALTinterwordspacing
OpenBCI. (2024) The complete ultracortex. [Online]. Available: \url{https://shop.openbci.com/products/the-complete-headset-eeg}
\BIBentrySTDinterwordspacing

\bibitem{Emotiv}
\BIBentryALTinterwordspacing
Emotiv. (2024) Epoc. [Online]. Available: \url{www.emotiv.com/epoc}
\BIBentrySTDinterwordspacing

\bibitem{veena2020review}
N.~Veena and N.~Anitha, ``A review of non-invasive bci devices,'' \emph{Int. J. Biomed. Eng. Technol}, vol.~34, no.~3, pp. 205--233, 2020.

\bibitem{Ali2020}
H.~A. Ali, N.~Goga, C.~V. Marian, and L.~A. Ali, ``An investigation of mind-controlled prosthetic arm intelligent system,'' \emph{eLearning and Software for Education Conference}, pp. 17--26, 2020.

\bibitem{Chinbat2018}
O.~Chinbat and J.~S. Lin, ``Prosthetic arm control by human brain,'' \emph{Proceedings - 2018 International Symposium on Computer, Consumer and Control, IS3C 2018}, pp. 54--57, 7 2018.

\bibitem{Beyrouthy2016}
T.~Beyrouthy, S.~K.~A. Kork, J.~A. Korbane, and A.~Abdulmonem, ``Eeg mind controlled smart prosthetic arm,'' \emph{2016 IEEE International Conference on Emerging Technologies and Innovative Business Practices for the Transformation of Societies, EmergiTech 2016}, pp. 404--409, 11 2016.

\bibitem{Lonsdale2020}
D.~Lonsdale, L.~I. Zhang, and R.~Jiang, ``3d printed brain-controlled robot-arm prosthetic via embedded deep learning from semg sensors,'' \emph{Proceedings - International Conference on Machine Learning and Cybernetics}, vol. 2020-December, pp. 247--253, 12 2020.

\bibitem{Zhang2024}
X.~Zhang, T.~Zhang, Y.~Jiang, W.~Zhang, Z.~Lu, Y.~Wang, and Q.~Tao, ``A novel brain-controlled prosthetic hand method integrating ar-ssvep augmentation, asynchronous control, and machine vision assistance,'' \emph{Heliyon}, vol.~10, p. e26521, 3 2024.

\bibitem{Vilela2020}
\BIBentryALTinterwordspacing
M.~Vilela and L.~R. Hochberg, ``Applications of brain-computer interfaces to the control of robotic and prosthetic arms,'' \emph{Handbook of clinical neurology}, vol. 168, pp. 87--99, 1 2020. [Online]. Available: \url{https://pubmed.ncbi.nlm.nih.gov/32164870/}
\BIBentrySTDinterwordspacing

\bibitem{MindArm}
\BIBentryALTinterwordspacing
M.~Nawaz, A.~Basit, and M.~Shafique, ``Mindarm: Mechanized intelligent non-invasive neuro-driven prosthetic arm system,'' 2024. [Online]. Available: \url{https://arxiv.org/abs/2403.19992}
\BIBentrySTDinterwordspacing

\bibitem{NeuroLimb}
A.~A. Cifuentes-Cuadros \emph{et~al.}, ``The libra neurolimb: Hybrid real-time control and mechatronic design for affordable prosthetics in developing regions,'' \emph{Sensors}, vol.~24, no.~1, 2024.

\bibitem{bionic}
\BIBentryALTinterwordspacing
Ottobock, ``Bebionic hand.'' [Online]. Available: \url{https://www.ottobock.com/en-us/product/8E7*}
\BIBentrySTDinterwordspacing

\bibitem{luke-arm}
\BIBentryALTinterwordspacing
Mobiusbionics, ``Luke arm.'' [Online]. Available: \url{https://mobiusbionics.com/luke-arm/}
\BIBentrySTDinterwordspacing

\bibitem{i-limb}
\BIBentryALTinterwordspacing
Ossur, ``i-ilimb.'' [Online]. Available: \url{https://www.ossur.com/en-us/prosthetics/arms/i-limb-quantum}
\BIBentrySTDinterwordspacing

\bibitem{Michelangelo}
\BIBentryALTinterwordspacing
Ottobock, ``Michelangelo hand.'' [Online]. Available: \url{https://www.ottobock.com/en-ex/product/8E500}
\BIBentrySTDinterwordspacing

\bibitem{ShadowHand}
S.~RObot, ``Shadow hand.''

\bibitem{Chen2023}
\BIBentryALTinterwordspacing
J.~Chen, Y.~Xia, X.~Zhou, E.~V. Rosas, A.~Thomas, R.~Loureiro, R.~J. Cooper, T.~Carlson, and H.~Zhao, ``fnirs-eeg bcis for motor rehabilitation: A review,'' \emph{Bioengineering 2023, Vol. 10, Page 1393}, vol.~10, p. 1393, 12 2023. [Online]. Available: \url{https://www.mdpi.com/2306-5354/10/12/1393/htm https://www.mdpi.com/2306-5354/10/12/1393}
\BIBentrySTDinterwordspacing

\bibitem{Kumari2023}
\BIBentryALTinterwordspacing
A.~Kumari and D.~R. Edla, ``A study on brain–computer interface: Methods and applications,'' \emph{SN Computer Science}, vol.~4, pp. 1--7, 3 2023. [Online]. Available: \url{https://link.springer.com/article/10.1007/s42979-022-01515-0}
\BIBentrySTDinterwordspacing

\bibitem{He2024}
\BIBentryALTinterwordspacing
Q.~He, Y.~Yang, P.~Ge, S.~Li, X.~Chai, Z.~Luo, and J.~Zhao, ``The brain nebula: minimally invasive brain–computer interface by endovascular neural recording and stimulation,'' \emph{Journal of NeuroInterventional Surgery}, vol.~0, pp. 1--7, 2 2024. [Online]. Available: \url{https://jnis.bmj.com/content/early/2024/02/22/jnis-2023-021296 https://jnis.bmj.com/content/early/2024/02/22/jnis-2023-021296.abstract}
\BIBentrySTDinterwordspacing

\bibitem{Abbady2022}
H.~E. Abbady \emph{et~al.}, ``3d-printed prostheses in developing countries: A systematic review,'' \emph{Prosthetics and Orthotics International}, vol.~46, pp. 19--30, 2 2022.

\bibitem{make3040042}
\BIBentryALTinterwordspacing
S.~Rasheed, ``A review of the role of machine learning techniques towards brain–computer interface applications,'' \emph{Machine Learning and Knowledge Extraction}, vol.~3, no.~4, pp. 835--862, 2021. [Online]. Available: \url{https://www.mdpi.com/2504-4990/3/4/42}
\BIBentrySTDinterwordspacing

\bibitem{brainflow}
\BIBentryALTinterwordspacing
A.~Parfenov \emph{et~al.}, ``Brainflow.'' [Online]. Available: \url{https://github.com/brainflow-dev/brainflow}
\BIBentrySTDinterwordspacing

\bibitem{Kothe2024}
C.~Kothe \emph{et~al.}, ``The lab streaming layer for synchronized multimodal recording,'' \emph{bioRxiv}, 2024.

\bibitem{shi2023flowformermaskedcostvolume}
\BIBentryALTinterwordspacing
X.~Shi, Z.~Huang \emph{et~al.}, ``Flowformer++: Masked cost volume autoencoding for pretraining optical flow estimation,'' 2023. [Online]. Available: \url{https://arxiv.org/abs/2303.01237}
\BIBentrySTDinterwordspacing

\bibitem{radford2022}
\BIBentryALTinterwordspacing
A.~Radford, J.~W. Kim, T.~Xu, G.~Brockman, C.~McLeavey, and I.~Sutskever, ``Robust speech recognition via large-scale weak supervision,'' 2022. [Online]. Available: \url{https://arxiv.org/abs/2212.04356}
\BIBentrySTDinterwordspacing

\end{thebibliography}
\endgroup

\end{document}